\documentclass[prd,twocolumn,twoside,preprintnumbers,superscriptaddress,nofootinbib]{revtex4}
\usepackage{amsmath,slashed}
\usepackage{graphicx,graphics,color}
\usepackage{dcolumn}
\usepackage[hyperfootnotes=false]{hyperref}
\usepackage{xspace}
\usepackage{enumerate}
\usepackage{varwidth}

\newcommand{\mpi}{M_\pi}

\renewcommand{\Im}{\text{Im}\,}
\renewcommand{\Re}{\text{Re}\,}
\newcommand{\Order}{\mathcal{O}}
\newcommand{\beq}{\begin{equation}}
\newcommand{\eeq}{\end{equation}}

\newcommand{\Li}{\text{Li}}

\allowdisplaybreaks[1]

\begin{document}

\preprint{LTH 1287}
\title{Mixed leptonic and hadronic corrections to the\\ anomalous magnetic moment of the muon}

\author{Martin Hoferichter}
\affiliation{Albert Einstein Center for Fundamental Physics, Institute for Theoretical Physics, University of Bern, Sidlerstrasse 5, 3012 Bern, Switzerland}
\author{Thomas Teubner}
\affiliation{Department of Mathematical Sciences, University of Liverpool, Liverpool, L69 3BX, U.K.}

\begin{abstract}
  Higher-order hadronic corrections to the anomalous magnetic moment of the muon have been evaluated including next-to-next-to-leading-order insertions of hadronic vacuum polarization and next-to-leading-order corrections to hadronic light-by-light scattering. This leaves a set of mixed leptonic and hadronic corrections in the form of double-bubble topologies as the only remaining hadronic effect at $\Order(\alpha^4)$. Here, we estimate these contributions by analyzing the respective cuts of the diagrams, suggesting that the impact is limited to $\lesssim 1\times 10^{-11}$ and thus negligible at the level of the final precision of the Fermilab $g-2$ experiment.        
\end{abstract}

\maketitle

\emph{Introduction}.---
The main uncertainty 
 in the Standard-Model prediction for the anomalous magnetic moment of the muon~\cite{Aoyama:2020ynm,Aoyama:2012wk,Aoyama:2019ryr,Czarnecki:2002nt,Gnendiger:2013pva,Davier:2017zfy,Keshavarzi:2018mgv,Colangelo:2018mtw,Hoferichter:2019gzf,Davier:2019can,Keshavarzi:2019abf,Hoid:2020xjs,Kurz:2014wya,Melnikov:2003xd,Colangelo:2014dfa,Colangelo:2014pva,Colangelo:2015ama,Masjuan:2017tvw,Colangelo:2017qdm,Colangelo:2017fiz,Hoferichter:2018dmo,Hoferichter:2018kwz,Gerardin:2019vio,Bijnens:2019ghy,Colangelo:2019lpu,Colangelo:2019uex,Blum:2019ugy,Colangelo:2014qya}
\beq
\label{amuSM}
a_\mu^\text{SM}=116\,591\,810(43)\times 10^{-11} 
\eeq
comes from the leading hadronic corrections: hadronic vacuum polarization (HVP) at $\Order(\alpha^2)$ in the expansion in the fine-structure constant $\alpha$ and hadronic light-by-light scattering (HLbL) at $\Order(\alpha^3)$. In addition, to achieve the required precision, higher-order corrections that include the insertion of the leading-order (LO) hadronic matrix elements have to be considered at next-to-leading order (NLO)~\cite{Calmet:1976kd} and even next-to-next-to-leading order (NNLO)~\cite{Kurz:2014wya}. Reference~\cite{Aoyama:2020ynm} includes the following contributions
\begin{align}
\label{hadronic}
 a_\mu^\text{HVP, LO}&=6\,931(40)\times 10^{-11} && \! \Order(\alpha^2)&& \!\! \text{\cite{Davier:2017zfy,Keshavarzi:2018mgv,Colangelo:2018mtw,Hoferichter:2019gzf,Davier:2019can,Keshavarzi:2019abf,Hoid:2020xjs}},\notag\\
 a_\mu^\text{HVP, NLO}&=-98.3(7)\times 10^{-11} && \! \Order(\alpha^3) && \!\! \text{\cite{Keshavarzi:2019abf}},\notag\\
  a_\mu^\text{HVP, NNLO}&=12.4(1)\times 10^{-11} && \! \Order(\alpha^4)&& \!\! \text{\cite{Kurz:2014wya}},\notag\\
  a_\mu^\text{HLbL}&=90(17)\times 10^{-11}&& \! \Order(\alpha^3)&& \!\! \text{\cite{Melnikov:2003xd,Colangelo:2014dfa,Colangelo:2014pva,Colangelo:2015ama,Masjuan:2017tvw,Colangelo:2017qdm,Colangelo:2017fiz,Hoferichter:2018dmo,Hoferichter:2018kwz,Gerardin:2019vio,Bijnens:2019ghy,Colangelo:2019lpu,Colangelo:2019uex,Pauk:2014rta,Danilkin:2016hnh,Jegerlehner:2017gek,Knecht:2018sci,Eichmann:2019bqf,Roig:2019reh,Blum:2019ugy}},\notag\\
  a_\mu^\text{HLbL, NLO}&=2(1)\times 10^{-11}&& \! \Order(\alpha^4)&& \!\! \text{\cite{Colangelo:2014qya}},
\end{align}
where we indicated the order in $\alpha$ at which each term arises.\footnote{We do not address possible tensions between lattice and data-driven determinations of the HVP contribution, see, e.g., Refs.~\cite{Borsanyi:2020mff,Lehner:2020crt,Crivellin:2020zul,Keshavarzi:2020bfy,Malaescu:2020zuc,Colangelo:2020lcg}. Note that for HLbL scattering there is good agreement between phenomenology and lattice QCD, see Refs.~\cite{Hoferichter:2020lap,Ludtke:2020moa,Bijnens:2020xnl,Bijnens:2021jqo,Zanke:2021wiq,Chao:2021tvp,Danilkin:2021icn,Colangelo:2021nkr} for some recent developments.}  
These numbers should be compared to the current experimental world average~\cite{bennett:2006fi,Abi:2021gix,Albahri:2021ixb,Albahri:2021kmg,Albahri:2021mtf}
\beq
\label{exp}
a_\mu^\text{exp}=116\,592\,061(41)\times 10^{-11},
\eeq
as well as the final precision $\Delta a_\mu^\text{exp}\text{[E989]}=16\times 10^{-11}$ projected for the Fermilab experiment~\cite{Muong-2:2015xgu}. Further experiments are planned at J-PARC~\cite{Abe:2019thb} and, potentially, at the proposed high-intensity-muon-beam facility at PSI~\cite{Aiba:2021bxe}, but in either case it appears challenging to move beyond a precision of $1\times 10^{-10}$. 

The comparison to Eq.~\eqref{hadronic} shows that while the uncertainties are well under control, even $\Order(\alpha^4)$ contributions do need to be included, given that the HVP contributions at NNLO are at the same level as $\Delta a_\mu^\text{exp}\text{[E989]}$. This surprising finding in Ref.~\cite{Kurz:2014wya} can be understood from enhancements that trace back to both large logarithms $\log\frac{m_\mu}{m_e}$ from electron loops and large numerical prefactors, as, e.g., expected from leptonic light-by-light topologies. The former also arises for NLO corrections to HLbL scattering, but in this case the corresponding enhancement does not counteract the suppression in $\alpha$ to the same extent~\cite{Colangelo:2014qya}. 

\begin{figure}[t]
	\centering
	\includegraphics[width=\linewidth,clip]{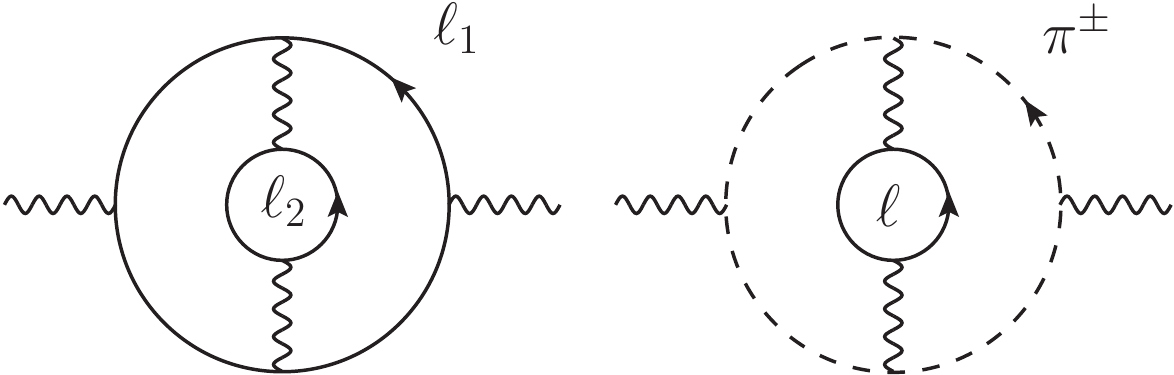}
	\caption{Left: double-bubble topology in QED, for outer lepton $\ell_1$ and inner lepton $\ell_2$. Right: example for a hadronic manifestation of the same topology with quarks in the outer loop and lepton $\ell$ in the inner loop. The opposite case (lepton outer loop, hadrons inner loop) is irrelevant numerically, see main text. Similar diagrams with inner-bubble insertions on the same side of the outer loop are not shown, nor are additional diagrams in scalar QED.}
	\label{fig:diagrams}
\end{figure}

In this Letter we address the remaining class of $\Order(\alpha^4)$ hadronic corrections, so-called double-bubble topologies shown in Fig.~\ref{fig:diagrams}. These contributions are subtle, in that care is required to avoid double counting with contributions that already enter LO HVP, given that purely hadronic cuts (and, possibly, to some extent mixed hadronic and leptonic cuts) are included in the measured $e^+e^-\to\text{hadrons}$ cross section.\footnote{We concentrate on the HVP determination from $e^+e^-$ data here. In lattice QCD, such topologies would require higher-order QED contributions.}
Numerically, potentially relevant effects are again expected from electron loops, and the subtleties in the definition of the LO HVP contribution further motivate a careful study of the different cuts to ensure that no sizable effects are overlooked. 
To this end, we first analyze the virtual (two-particle cut) and real (four-particle cut) contributions in QED, following the methods developed in the context of higher-order corrections to heavy-quark production~\cite{Teubner:1995,Hoang:1995,Hoang:1993ns,Hoang:1994it,Hoang:1995ex,Hoang:1995ht,Brodsky:1995ds,Chetyrkin:1996yp,Hoang:1997ca}, and then generalize the results to scalar QED to estimate the corrections originating from the leading hadronic channel $e^+e^-\to\pi^+\pi^-$. This strategy allows us to calculate complicated four-loop contributions in an efficient and transparent manner, since the spectral functions that emerge at intermediate steps directly correspond to physical cross sections.     

\emph{QED}.---The vacuum polarization (VP) diagrams in Fig.~\ref{fig:diagrams} produce a contribution to $a_\mu$ via
\beq
\label{amu}
a_\mu=\frac{\alpha}{\pi}\int_0^1dx (1-x)\bar\Pi(s_x),\quad s_x=-\frac{x^2m_\mu^2}{1-x},
\eeq
where $\bar\Pi(s)$ is the renormalized scalar VP function in the sign convention that the fine-structure constant runs as $\alpha(s)=\alpha(0)/(1-\bar\Pi(s))$. We will evaluate Eq.~\eqref{amu} via the dispersion relation
\beq
\label{Pibardisp}
\bar\Pi(s)=\frac{s}{\pi}\int_{s_\text{thr}}^\infty ds' \frac{\Im \Pi(s')}{s'(s'-s)},
\eeq
since $\Im \Pi(s)$ can be directly related to the cuts of the diagrams. To be explicit, one has the relation 
\beq
\Im \Pi(s)=-\frac{\alpha}{3}R(s),
\eeq
where the $R$-ratio is defined as
\beq
R(s)=\frac{\sigma(e^+e^-\to\text{hadrons})}{\sigma_0},\qquad \sigma_0=\frac{4\pi\alpha^2}{3s}.
\eeq
Of course, the same formula also works for leptonic final states, so that the left diagram in Fig.~\ref{fig:diagrams} can be reconstructed from the $\ell_1^+\ell_1^-$ cut, starting at $s_\text{thr}=4m_{\ell_1}^2$, and the $\ell_1^+\ell_1^-\ell_2^+\ell_2^-$ cut, starting at $s_\text{thr}=(2m_{\ell_1}+2m_{\ell_2})^2$. As a first step, we work out the results for the cases $\{\ell_1\ell_2\}=\{ee,\mu e,e\mu\}$, since the separation into the two cuts, referred to as virtual and real contributions, respectively, will allow us to draw first conclusions on the hadronic case. 

To this end, we introduce the notation 
\beq
R^{(2)}_{\ell_1\ell_2}(s)=\bigg(\frac{\alpha}{\pi}\bigg)^2\big(\rho^V_{\ell_1\ell_2}(s)+\rho^R_{\ell_1\ell_2}(s)\big)
\eeq
for the two-loop contribution to $R(s)$ and separate the scaling in $\alpha/\pi$ to obtain the spectral functions $\rho^{V,R}_{\ell_1\ell_2}$ for the virtual and real parts. 

The virtual spectral function can be calculated by yet another dispersion relation,
\begin{align}
\label{rhoVlambda}
 \rho^{V}_{\ell_1\ell_2}(s)&=(3-\beta_1^2)\beta_1 \Re F_1^{\ell_1\ell_2}(s) + 3\beta_1\Re F_2^{\ell_1\ell_2}(s),\notag\\
 F_i^{\ell_1\ell_2}(s)&=\frac{1}{3}\int_{4m_2^2}^\infty \frac{d\lambda^2}{\lambda^2}F_i^{\ell_1\lambda}(s,\lambda^2) R_{\ell_2}(\lambda^2),
\end{align}
where 
\beq
R_\ell(s)=\sqrt{1-\frac{4m_{\ell}^2}{s}}\bigg(1+\frac{2m_{\ell}^2}{s}\bigg)
\eeq
is the spectral function for the inner lepton $\ell_2$
and we wrote $m_i=m_{\ell_i}$ as well as $\beta_i=\sqrt{1-4m_i^2/s}$. The $F_i^{\ell_1\lambda}(s,\lambda^2)$ are the one-loop QED Dirac and Pauli form factors, but calculated for a finite photon mass $\lambda$~\cite{Hoang:1995ex,Hoang:1997ca}, see Eq.~\eqref{FilambdaQED} 
 for the explicit expressions~\cite{appendix}. The resulting integral representation~\eqref{rhoVlambda} works well for $\ell_1=\mu$, leading to the result shown in Table~\ref{tab:numerics}. For $\ell_1=\ell_2=e$ one can use the analytic result~\cite{Teubner:1995,Hoang:1995}, reproduced in Eq.~\eqref{rhoV_equal}, while for $\ell_1=e$, $\ell_2=\mu$ in most of the integration range in Eq.~\eqref{Pibardisp} the approximation $m_1=0$ is sufficient. The exception is the region very close to threshold, where a double expansion~\eqref{rhoVm1exp} in
 $\beta_1$ and $m_1^2/\lambda^2$ should be used instead. Finally, the known real spectral function 
 from the four-particle cut is given in Eq.~\eqref{rhoR}.

\begin{table}[t]
\renewcommand{\arraystretch}{1.3}
	\centering
	\begin{tabular}{l r r r r}
	\toprule
	& virtual & real & total & Refs.~\cite{Aoyama:2012wk,Nio:2021}\\\colrule
	$\ell_1=\ell_2=e$& $-3.56$& $5.00$ & $1.44$& $1.4407623(330)$\\
	$\ell_1=\mu, \ell_2=e$ & $-0.11$ & $0.27$ & $0.16$ & $0.1619701(214)$\\
	$\ell_1=e, \ell_2=\mu$ & $0.02$ & $0.002$ & $0.02$ & $0.0215813(40)$\\\colrule
	$\ell=e$ & $-0.13$ & $0.25$ & $0.12$ &\\
\botrule
	\end{tabular}
	\caption{Contributions to $a_\mu$ in units of $(\alpha/\pi)^4\simeq 2.9\times 10^{-11}$. The upper panel refers to the QED diagrams (left in Fig.~\ref{fig:diagrams}), the last line to an estimate of the $\pi^+\pi^-$ contribution in scalar QED (right in Fig.~\ref{fig:diagrams}). The virtual and real parts correspond to the two- and four-particle cuts, respectively.}
	\label{tab:numerics}
\end{table}

The sum of the real and virtual contributions reproduces the total results from Refs.~\cite{Aoyama:2012wk,Nio:2021}, see Table~\ref{tab:numerics}.\footnote{Related work on the mass-dependent $\Order(\alpha^4)$ QED contributions can be found in Refs.~\cite{Laporta:1993ds,Kinoshita:2005zr,Kataev:2012kn,Kurz:2013exa,Kurz:2015bia,Kurz:2016bau}.} We see that by far the dominant contribution arises from $\ell_1=\ell_2=e$, as expected given the additional light loop compared to the other two cases. Moreover, we observe that (i) the configuration with an outer electron and inner muon comes out much smaller than the opposite case and (ii) the cancellation of the leading logarithms is most visible for $\ell_1=\ell_2=e$, with a milder effect already for $\ell_1=\mu$, $\ell_2=e$.     

Since the muon mass is similar to the hadronic scales, e.g., close to the mass of the pion, these observations provide some first insights for the corresponding diagrams with hadronic degrees of freedom. First, we can ignore the case of on outer electron and inner hadronic loop, since the muon example shows that this configuration contributes only $\lesssim 10^{-12}$ to $a_\mu$. Second, virtual corrections will be included in the $e^+e^-\to\text{hadrons}$ data in LO HVP, unless removed by hand through the application of higher-order radiative corrections or in Monte Carlo simulations. This implies that the only potentially missing effect concerns the real radiation of an $e^+e^-$ pair together with hadronic states. In the muon case, this effect amounts to $\Delta a_\mu\simeq 0.8\times 10^{-11}$, which would be negligible for the time being. To corroborate this estimate, we consider a $\pi^+\pi^-$ loop in scalar QED, as a realistic example of the hadronic realization of a quark loop, see right diagram in Fig.~\ref{fig:diagrams}.

\emph{Scalar QED}.---To estimate the potentially missing hadronic contributions  more quantitatively, we consider the $e^+e^-\to\pi^+\pi^-$ cross section parameterized via the pion vector form factor $F_\pi^V(s)$,
\beq
\sigma(e^+e^-\to\pi^+\pi^-)=\frac{\pi\alpha^2}{3s}\beta_\pi^3 \big|F_\pi^V(s)\big|^2,
\eeq
with $\beta_\pi=\sqrt{1-4\mpi^2/s}$, 
and evaluate the virtual and real corrections in scalar QED. This strategy is analogous to the calculation of $\pi\pi\gamma$ radiative corrections~\cite{Hoefer:2001mx,Czyz:2004rj,Gluza:2002ui,Bystritskiy:2005ib} and captures the infrared enhanced effects, for which the pion can be approximated as a point-like particle. 

The calculation of the virtual contribution proceeds in analogy to Eq.~\eqref{rhoVlambda},
\begin{align}
\rho_{\pi\ell}^V(s)&=\frac{1}{2}\beta_\pi^3\big|F_\pi^V(s)\big|^2 \Re F^{\pi\ell}(s),\notag\\
F^{\pi\ell}(s)&=\frac{1}{3}\int_{4m_\ell^2}^\infty \frac{d\lambda^2}{\lambda^2}F^{\pi\lambda}(s,\lambda^2) R_{\ell}(\lambda^2),
\end{align}
where 
the form factor $F^{\pi\lambda}(s,\lambda^2)$ fulfills the dispersion relation
\beq
F^{\pi\lambda}(s,\lambda^2)=\frac{s}{\pi}\int_{4\mpi^2}^\infty ds'\frac{\Im F^{\pi\lambda}(s',\lambda^2)}{s'(s'-s)}.
\eeq
From the explicit scalar QED calculation we obtain the compact analytic expression
\begin{widetext}
 \begin{align}
 \label{Fpilambda}
F^{\pi\lambda}(s,\lambda^2)&=-\frac{1}{4\beta}\bigg[1+\beta^2+\frac{2+\beta^2-3\beta^4}{4\beta^2}l+\frac{(1-\beta^2)^2}{8\beta^2}l^2\bigg]\psi(p,l)-\bigg[\frac{1}{2\beta^2}+\frac{1+\beta^2}{8\beta^2}l-\frac{1}{4}\bigg]\phi(l)\notag\\
&-\frac{1}{16\beta^2}\bigg[1+\beta^2+\frac{(1-\beta^2)^2}{2\beta}\log p\bigg]l^2\log l
-\frac{1}{16\beta^2}\bigg[4-6\beta^2+\frac{2+\beta^2-3\beta^4}{\beta}\log p\bigg]l\log l-\frac{(1-\beta^2)^2\log^2 p}{64\beta^3}l^2\notag\\
&+\frac{1}{4}\bigg[1-\frac{1-\beta^2}{8\beta^3}\log p\Big(4\beta^2+(2+3\beta^2)\log p\Big)\bigg]l
-\frac{1}{2}\bigg[1+\frac{1+\beta^2}{2\beta}\log p\bigg]\log l-\frac{\log p}{8\beta}(1+\beta^2)(4+\log p)-1\notag\\
&+i\pi\Bigg\{-\frac{1}{2\beta}\bigg[1+\beta^2+\frac{1-\beta^2}{4}l\bigg]+\frac{1}{4\beta}\bigg[1+\beta^2+\frac{2+\beta^2-3\beta^4}{4\beta^2}l+\frac{(1-\beta^2)^2}{8\beta^2}l^2\bigg]\log\bigg[1+\frac{(1-p)^2}{l p}\bigg]\Bigg\},
\end{align}
where $\beta=\beta_\pi$, $p=\frac{1-\beta}{1+\beta}$, $l=\lambda^2/\mpi^2$, and
\begin{align}
\label{phi_psi}
 \phi(l)&=\begin{cases}
           \frac{1}{2}\sqrt{l^2-4l}\log\frac{l-\sqrt{l^2-4l}}{l+\sqrt{l^2-4l}}\qquad \text{for}\quad l>4\\
           \sqrt{4l-l^2}\arctan\frac{\sqrt{4l-l^2}}{l}\qquad \text{for} \quad 0 < l < 4
          \end{cases},\notag\\
          \psi(p,l)&=\frac{1}{2}\log^2\bigg(\frac{1}{2}\Big[l-2+\sqrt{l^2-4l}\Big]\bigg)+\Li_2\bigg(1+\frac{p}{2}\Big[-2+l+\sqrt{l^2-4l}\Big]\bigg)+\Li_2\bigg(1+\frac{p}{2}\Big[-2+l-\sqrt{l^2-4l}\Big]\bigg)\notag\\
          &+\begin{cases}
             -\frac{\pi^2}{2}+i\pi\log\Big((1-p)^2+p l\Big)\qquad \text{for}\quad l>4\\
             -\frac{3}{2}\pi^2+4\pi\arctan\frac{\sqrt{4l-l^2}}{l}+2\pi\arctan\frac{2p+l-2}{\sqrt{4l-l^2}} \qquad \text{for} \quad 0 < l < 4
            \end{cases}.
\end{align}
Similarly, real radiation from the scalar QED subprocess leads to the spectral function 
\begin{align}
\label{rhoRpion}
\rho^R_{\pi\ell}(s)&=\frac{4}{3}\big|F_\pi^V(s)\big|^2\int_{y_-}^{y_+}dy\int_{z_-}^{z_+} \frac{dz}{z}\bigg(1+\frac{2m_\ell^2}{s z}\bigg)\sqrt{1-\frac{4m_\ell^2}{sz}}\Bigg[ \sqrt{1-\frac{4\mpi^2}{s y}}\lambda^{1/2}(1,y,z)\Bigg(\frac{1}{8}+\frac{\Big(\frac{4\mpi^2}{s}-1\Big)\Big(\frac{4\mpi^2}{s}-z\Big)}{32\big(z+\frac{\mpi^2}{sy}\lambda(1,y,z)\big)}\Bigg)\notag\\
 &+\frac{\frac{16\mpi^4}{s^2}-\frac{4\mpi^2}{s}(2y+z+1)+2y(z+1)+z}{16(y-z-1)}\log\frac{1-y+z-\sqrt{1-\frac{4\mpi^2}{s y}}\lambda^{1/2}(1,y,z)}{1-y+z+\sqrt{1-\frac{4\mpi^2}{s y}}\lambda^{1/2}(1,y,z)}
\Bigg],
\end{align}
\end{widetext}
where $\lambda(x,y,z)=x^2+y^2+z^2-2(x y+x z+y z)$ and
\begin{align}
\label{limits}
y_-&=\frac{4\mpi^2}{s}, & y_+&=\bigg(1-\frac{2m_\ell}{\sqrt{s}}\bigg)^2,\notag\\
z_-&=\frac{4m_\ell^2}{s}, & z_+&=\big(1-\sqrt{y}\big)^2.
\end{align}
Both $F^{\pi\lambda}(s,\lambda^2)$ and $\rho^R_{\pi \ell}(s)$ are new results.

\begin{figure}[t!]
	\centering
	\includegraphics[width=\linewidth,clip]{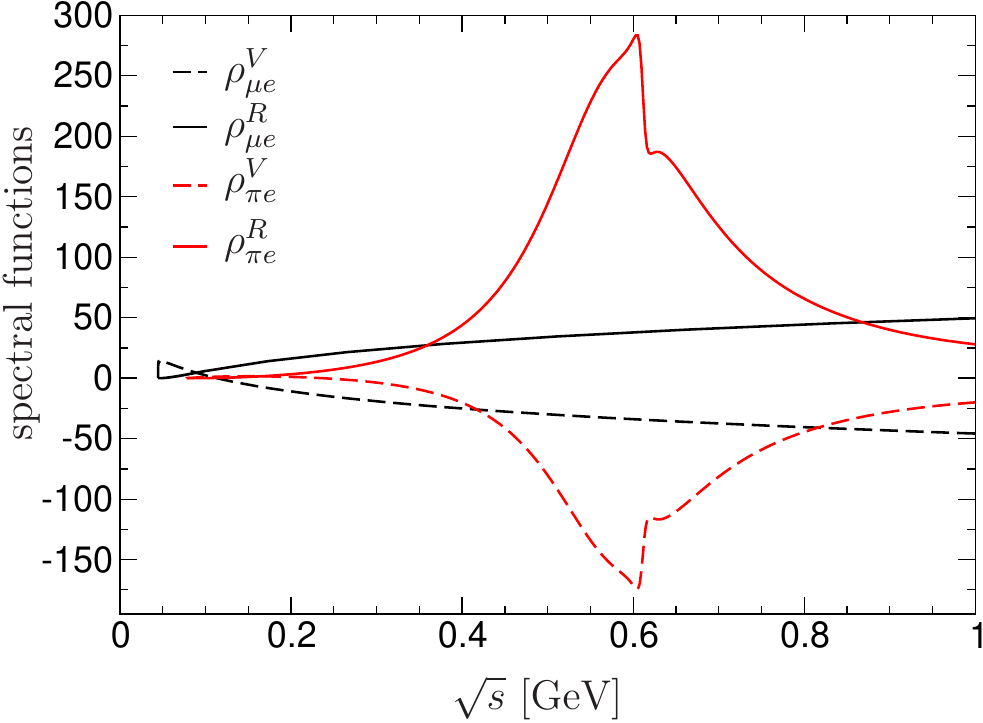}
	\caption{Spectral functions for an inner electron and an outer muon or pion. The figure illustrates how the larger fermionic spectral functions close to threshold and for large $s$ are compensated by the $\rho(770)$ peak in between.}
	\label{fig:spectral_functions}
\end{figure}

The numerical evaluation gives the result in the last line of Table~\ref{tab:numerics}, supporting the conclusions already indicated by the muon example: again, the contribution from the real radiation of $e^+e^-$ pairs together with final-state $\pi^+\pi^-$ corresponds to an effect in $a_\mu$ of less than $1\times 10^{-11}$. We stress that, while the muon loop is expected to produce results similar to hadronic degrees of freedom given the scales involved, the extent to which the effects in $a_\mu$ agree is largely coincidental. As shown in Fig.~\ref{fig:spectral_functions}, the fermionic spectral functions tend to be larger near threshold (due to the $P$-wave suppression of the $\pi^+\pi^-$ channel) and for large $s$ (where $F_\pi^V\simeq 1/s$ leads to a suppression), which is then compensated by the $\rho(770)$ peak at intermediate energies.

\emph{Conclusions}.---In this Letter we addressed a missing class of hadronic corrections to the anomalous magnetic moment of the muon at $\Order(\alpha^4)$, represented by the double-bubble topologies shown in Fig.~\ref{fig:diagrams}. As a first step we reproduced the QED configurations involving electron loops by means of the two- and four-particle cuts of these diagrams, and showed that indeed the known results are recovered when adding the virtual and real contributions. In particular, configurations in which the inner loop does not correspond to an electron prove negligible.   

Since virtual corrections should be included in the measured $e^+e^-\to\text{hadrons}$ cross section, we concluded that the only potentially missed contribution could originate from
hadronic final states accompanied by $e^+e^-$ pair emission. To estimate this effect, we argued that the case with an outer muon should provide a first indication, and then calculated the analog in scalar QED for a charged-pion loop. Both results are remarkably close, albeit to a large extent by coincidence, see Fig.~\ref{fig:spectral_functions}, and translate to an effect in $a_\mu$ of $\lesssim 1\times 10^{-11}$. One would thus need a significant enhancement due to experimental cuts or misidentification of $e^+e^-$ pairs to produce a relevant effect. It might be interesting to verify in experimental analyses that indeed no enhanced effects from $e^+e^-$ radiation can occur, but absent such a scenario we 
 conclude that this class of mixed leptonic and hadronic corrections is negligible at the level required for the final precision of the Fermilab $g-2$ experiment.

\emph{Acknowledgments}.---We thank Makiko Nio for providing the flavor breakdown of the results from Ref.~\cite{Aoyama:2012wk}, and acknowledge valuable discussions at the workshop~\cite{Abbiendi:2022liz}.
Financial support by  the SNSF under Project No.\ PCEFP2\_181117 (M.H.) and the STFC Consolidated
Grant ST/T000988/1 (T.T.) is gratefully acknowledged.

\appendix

\begin{widetext}

\section{QED spectral functions}

The QED form factors for finite $\lambda^2$ are
\begin{align}
 \label{FilambdaQED}
 F_1^{\ell\lambda}&=-\frac{1}{4\beta}\bigg[1+\beta^2+\frac{1-\beta^2}{\beta^2}l\bigg(1+\frac{(3-2\beta^2)(1-\beta^2)}{8\beta^2} l\bigg)\bigg]\psi(p,l)
 +\bigg[\frac{-1+2\beta^2}{2\beta^2}+\frac{-3+7\beta^2+2\beta^4}{8\beta^4}l+\frac{3}{l-4}\bigg]\phi(l)\notag\\
 &+\frac{1}{16\beta^4}\bigg[-3+7\beta^2+2\beta^4+\frac{(1-\beta^2)^2(-3+2\beta^2)}{2\beta}\log p\bigg]l^2\log l-\frac{1}{2\beta^2}\bigg[1+\frac{1-\beta^2}{2\beta}\log p\bigg] l\log l\notag\\
 &-\bigg[\frac{1}{2}+\frac{1+\beta^2}{4\beta}\log p\bigg]\log l
 +\frac{(1-\beta^2)^2(-3+2\beta^2)\log^2p}{64\beta^5} l^2-\frac{1}{4\beta^2}\bigg[1+2\beta^2+\frac{1-\beta^2}{2\beta}\log p\Big(3-2\beta^2+\log p\Big)\bigg]l\notag\\
 &-\frac{\log p}{8\beta}\bigg[2(1+2\beta^2)+(1+\beta^2)\log p\bigg]-1\notag\\
 &+i\pi\Bigg\{-\frac{1+2\beta^2}{4\beta}+\frac{(1-\beta^2)(-3+2\beta^2)}{8\beta^3}l+\frac{1}{4\beta}\bigg[1+\beta^2+\frac{1-\beta^2}{\beta^2}l+\frac{(3-2\beta^2)(1-\beta^2)^2}{8\beta^4}l^2\bigg]\log\bigg[1+\frac{(1-p)^2}{l p}\bigg]\Bigg\},\notag\\
 F_2^{\ell\lambda}&=\frac{(1-\beta^2)^2}{4\beta^3}\bigg[1+\frac{3(1-\beta^2)}{8\beta^2}l\bigg]l\,\psi(p,l)+\frac{1-\beta^2}{2\beta^2}\bigg[1+\frac{3-5\beta^2}{4\beta^2}l\bigg]\phi(l)\notag\\
 &+\frac{(1+\beta)^2(1-\beta^2)}{16\beta^4}\bigg[\frac{3-5\beta^2}{(1+\beta)^2}+\frac{3(1-\beta)^2}{2\beta}\log p\bigg]l^2\log l
 +\frac{1-\beta^2}{2\beta^2}\bigg[1+\frac{1-\beta^2}{2\beta}\log p\bigg]l\log l+\frac{3(1-\beta^2)^3}{64\beta^5}l^2 \log^2p\notag\\
 &+\frac{1-\beta^2}{4\beta^2}\bigg[1+\frac{1-\beta^2}{2\beta}\log p(3+\log p)\bigg]l+\frac{1-\beta^2}{4\beta}\log p\notag\\
 &+i\pi\Bigg\{\frac{1-\beta^2}{4\beta}+\frac{3(1-\beta^2)^2}{8\beta^3}l
 -\frac{(1-\beta^2)^2}{4\beta^3}\bigg[1+\frac{3(1-\beta^2)}{8\beta^2}l\bigg]l\,\log\bigg[1+\frac{(1-p)^2}{l p}\bigg]\Bigg\},
\end{align}
where $\beta=\sqrt{1-\frac{4m_\ell^2}{s}}$, $p=\frac{1-\beta}{1+\beta}$, $l=\lambda^2/m_\ell^2$, and $\phi(l)$, $\psi(p,l)$ as in Eq.~\eqref{phi_psi}. 
In the equal-mass case $\ell_1=\ell_2=\ell$ also compact expressions for the integral~\eqref{rhoVlambda} exist~\cite{Teubner:1995,Hoang:1995},
\begin{align}
\label{rhoV_equal}
 \Re F_1^{\ell\ell}&=\frac{1}{12\beta^2}\Bigg[\frac{(2 \beta
   ^2-1) (9-6 \beta ^2+5 \beta ^4) }{24 \beta
   ^3}\log^3p+\frac{31-23 \beta ^2+30 \beta ^4}{12 \beta ^2}\log^2 p\\
   &+\bigg(\frac{267-238 \beta ^2+236 \beta ^4}{18 \beta }+\frac{(1-2 \beta
   ^2) (9-6 \beta ^2+5 \beta ^4) \zeta(2)}{2 \beta
   ^3}\bigg)\log p+\frac{(9-21 \beta ^2-10 \beta ^4) \zeta(2)}{\beta
   ^2}+\frac{147+236 \beta ^2}{9}\Bigg],\notag\\
\Re F_2^{\ell\ell}&=\frac{1-\beta ^2}{4\beta^2}\Bigg[ \frac{(1-\beta
   ^2)^2}{8 \beta ^3}\log^3p-\frac{11-9 \beta ^2}{12 \beta ^2}\log^2p-\bigg(\frac{93-68 \beta ^2}{18 \beta }+\frac{3
   (1-\beta ^2)^2 \zeta(2)}{2 \beta ^3}\bigg)\log p+\frac{(5 \beta
   ^2-3) \zeta(2)}{\beta ^2}-\frac{17}{3}\Bigg].\notag
\end{align}
 In the case $\ell_1=e$, $\ell_2=\mu$ the approximation $m_1=0$ is useful for most of the integration range, leading to
\begin{align}
\label{rhoVm1zero}
 \rho^V_{\ell_1\ell_2}\big|_{m_1=0}&=\frac{2}{3}\big(1-6x^2\big)\Big[\Li_3\big(A^2\big)-\zeta(3)-2\zeta(2)\log A+\frac{2}{3}\log^3A\Big]+\frac{19+46x}{9}\sqrt{1+4x}\Big[\Li_2\big(A^2\big)-\zeta(2)+\log^2A\Big]
 \notag\\
 &+\frac{5}{36}\bigg(\frac{53}{3}+44x\bigg)\log x+\frac{3355}{648}+\frac{119}{9}x,\qquad A=\frac{\sqrt{1+4x}-1}{\sqrt{4x}},\qquad x=\frac{m_2^2}{s},
\end{align}
while close to threshold the following double expansion in $\beta=\beta_1$ and $l=\lambda^2/m_1^2$ applies:
\begin{align}
 \label{rhoVm1exp}
 \Re F_1^{\ell\lambda}&=\frac{1}{10395l}\bigg[19635-8085\beta^2-8085\beta^4-9933\beta^6-11517\beta^8-12837\beta^{10}-13957\beta^{12}+6930\log l\,\sum\limits_{i=0}^6 \beta^{2i}\bigg]\notag\\
 &+\frac{1}{3780l^2}\bigg[
 525(29-5\beta^2)+14805\beta^4+23835\beta^6+31857\beta^8+39735\beta^{10}+47693\beta^{12}-1260\log l\,\sum_{i=0}^6(2i+5)\beta^{2i}\bigg]\notag\\
 &+\Order\Big(\beta^{14}l^{-1}\log l,l^{-3}\log l\Big),\notag\\
 &\notag\\
 \Re F_2^{\ell\lambda}&=\frac{1}{3780l^2}\bigg[1785-1050\beta^2-6090\beta^4-7770\beta^6-8778\beta^8-9498\beta^{10}-10058\beta^{12}-1260\log l+2520\log l\,\sum_{i=1}^6\beta^{2i}\bigg]\notag\\
&+\frac{1}{3l}+\Order\Big(\beta^{14}l^{-2}\log l,l^{-3}\log l\Big).
\end{align} 
The four-particle cut leads to the spectral function
\begin{align}
 \label{rhoR}
 \rho^R_{\ell_1\ell_2}&=\frac{4}{3}\int_{y_-}^{y_+}dy\int_{z_-}^{z_+} \frac{dz}{z}\bigg(1+\frac{2m_2^2}{s z}\bigg)\sqrt{1-\frac{4m_2^2}{sz}}\Bigg[\sqrt{1-\frac{4m_1^2}{s y}}\lambda^{1/2}(1,y,z)\Bigg(\frac{\frac{2m_1^2}{s}+\frac{4m_1^4}{s^2}+\Big(1+\frac{2m_1^2}{s}\Big)z}{\Big(1-\frac{4m_1^2}{s y}\Big)\lambda(1,y,z)-(1-y+z)^2}-\frac{1}{4}\Bigg)\notag\\
 &+\frac{\frac{2m_1^4}{s^2}+\frac{m_1^2}{s}(1-y+z)-\frac{1}{4}(1-y+z)^2-\frac{1}{2}(1+z)y}{1-y+z}\log\frac{1-y+z-\sqrt{1-\frac{4m_1^2}{s y}}\lambda^{1/2}(1,y,z)}{1-y+z+\sqrt{1-\frac{4m_1^2}{s y}}\lambda^{1/2}(1,y,z)}
\Bigg],
\end{align}
where
\beq
\label{limits}
y_-=\frac{4m_1^2}{s},\qquad y_+=\bigg(1-\frac{2m_2}{\sqrt{s}}\bigg)^2,\qquad 
z_-=\frac{4m_2^2}{s},\qquad z_+=\big(1-\sqrt{y}\big)^2.
\eeq
\end{widetext}

\bibliography{AMM}

\begin{thebibliography}{79}
\expandafter\ifx\csname natexlab\endcsname\relax\def\natexlab#1{#1}\fi
\expandafter\ifx\csname bibnamefont\endcsname\relax
  \def\bibnamefont#1{#1}\fi
\expandafter\ifx\csname bibfnamefont\endcsname\relax
  \def\bibfnamefont#1{#1}\fi
\expandafter\ifx\csname citenamefont\endcsname\relax
  \def\citenamefont#1{#1}\fi
\expandafter\ifx\csname url\endcsname\relax
  \def\url#1{\texttt{#1}}\fi
\expandafter\ifx\csname urlprefix\endcsname\relax\def\urlprefix{URL }\fi
\providecommand{\bibinfo}[2]{#2}
\providecommand{\eprint}[2][]{\url{#2}}

\bibitem[{\citenamefont{Aoyama et~al.}(2020)}]{Aoyama:2020ynm}
\bibinfo{author}{\bibfnamefont{T.}~\bibnamefont{Aoyama}} \bibnamefont{et~al.},
  \bibinfo{journal}{Phys. Rept.} \textbf{\bibinfo{volume}{887}},
  \bibinfo{pages}{1} (\bibinfo{year}{2020}), \eprint{2006.04822}.

\bibitem[{\citenamefont{Aoyama et~al.}(2012)\citenamefont{Aoyama, Hayakawa,
  Kinoshita, and Nio}}]{Aoyama:2012wk}
\bibinfo{author}{\bibfnamefont{T.}~\bibnamefont{Aoyama}},
  \bibinfo{author}{\bibfnamefont{M.}~\bibnamefont{Hayakawa}},
  \bibinfo{author}{\bibfnamefont{T.}~\bibnamefont{Kinoshita}},
  \bibnamefont{and} \bibinfo{author}{\bibfnamefont{M.}~\bibnamefont{Nio}},
  \bibinfo{journal}{Phys. Rev. Lett.} \textbf{\bibinfo{volume}{109}},
  \bibinfo{pages}{111808} (\bibinfo{year}{2012}), \eprint{1205.5370}.

\bibitem[{\citenamefont{Aoyama et~al.}(2019)\citenamefont{Aoyama, Kinoshita,
  and Nio}}]{Aoyama:2019ryr}
\bibinfo{author}{\bibfnamefont{T.}~\bibnamefont{Aoyama}},
  \bibinfo{author}{\bibfnamefont{T.}~\bibnamefont{Kinoshita}},
  \bibnamefont{and} \bibinfo{author}{\bibfnamefont{M.}~\bibnamefont{Nio}},
  \bibinfo{journal}{Atoms} \textbf{\bibinfo{volume}{7}}, \bibinfo{pages}{28}
  (\bibinfo{year}{2019}).

\bibitem[{\citenamefont{Czarnecki et~al.}(2003)\citenamefont{Czarnecki,
  Marciano, and Vainshtein}}]{Czarnecki:2002nt}
\bibinfo{author}{\bibfnamefont{A.}~\bibnamefont{Czarnecki}},
  \bibinfo{author}{\bibfnamefont{W.~J.} \bibnamefont{Marciano}},
  \bibnamefont{and}
  \bibinfo{author}{\bibfnamefont{A.}~\bibnamefont{Vainshtein}},
  \bibinfo{journal}{Phys. Rev. D} \textbf{\bibinfo{volume}{67}},
  \bibinfo{pages}{073006} (\bibinfo{year}{2003}), \bibinfo{note}{[Erratum:
  Phys. Rev. D {\bf 73}, 119901 (2006)]}, \eprint{hep-ph/0212229}.

\bibitem[{\citenamefont{Gnendiger et~al.}(2013)\citenamefont{Gnendiger,
  St{\"o}ckinger, and St{\"o}ckinger-Kim}}]{Gnendiger:2013pva}
\bibinfo{author}{\bibfnamefont{C.}~\bibnamefont{Gnendiger}},
  \bibinfo{author}{\bibfnamefont{D.}~\bibnamefont{St{\"o}ckinger}},
  \bibnamefont{and}
  \bibinfo{author}{\bibfnamefont{H.}~\bibnamefont{St{\"o}ckinger-Kim}},
  \bibinfo{journal}{Phys. Rev. D} \textbf{\bibinfo{volume}{88}},
  \bibinfo{pages}{053005} (\bibinfo{year}{2013}), \eprint{1306.5546}.

\bibitem[{\citenamefont{Davier et~al.}(2017)\citenamefont{Davier, Hoecker,
  Malaescu, and Zhang}}]{Davier:2017zfy}
\bibinfo{author}{\bibfnamefont{M.}~\bibnamefont{Davier}},
  \bibinfo{author}{\bibfnamefont{A.}~\bibnamefont{Hoecker}},
  \bibinfo{author}{\bibfnamefont{B.}~\bibnamefont{Malaescu}}, \bibnamefont{and}
  \bibinfo{author}{\bibfnamefont{Z.}~\bibnamefont{Zhang}},
  \bibinfo{journal}{Eur. Phys. J. C} \textbf{\bibinfo{volume}{77}},
  \bibinfo{pages}{827} (\bibinfo{year}{2017}), \eprint{1706.09436}.

\bibitem[{\citenamefont{Keshavarzi et~al.}(2018)\citenamefont{Keshavarzi,
  Nomura, and Teubner}}]{Keshavarzi:2018mgv}
\bibinfo{author}{\bibfnamefont{A.}~\bibnamefont{Keshavarzi}},
  \bibinfo{author}{\bibfnamefont{D.}~\bibnamefont{Nomura}}, \bibnamefont{and}
  \bibinfo{author}{\bibfnamefont{T.}~\bibnamefont{Teubner}},
  \bibinfo{journal}{Phys. Rev. D} \textbf{\bibinfo{volume}{97}},
  \bibinfo{pages}{114025} (\bibinfo{year}{2018}), \eprint{1802.02995}.

\bibitem[{\citenamefont{Colangelo et~al.}(2019)\citenamefont{Colangelo,
  Hoferichter, and Stoffer}}]{Colangelo:2018mtw}
\bibinfo{author}{\bibfnamefont{G.}~\bibnamefont{Colangelo}},
  \bibinfo{author}{\bibfnamefont{M.}~\bibnamefont{Hoferichter}},
  \bibnamefont{and} \bibinfo{author}{\bibfnamefont{P.}~\bibnamefont{Stoffer}},
  \bibinfo{journal}{JHEP} \textbf{\bibinfo{volume}{02}}, \bibinfo{pages}{006}
  (\bibinfo{year}{2019}), \eprint{1810.00007}.

\bibitem[{\citenamefont{Hoferichter et~al.}(2019)\citenamefont{Hoferichter,
  Hoid, and Kubis}}]{Hoferichter:2019gzf}
\bibinfo{author}{\bibfnamefont{M.}~\bibnamefont{Hoferichter}},
  \bibinfo{author}{\bibfnamefont{B.-L.} \bibnamefont{Hoid}}, \bibnamefont{and}
  \bibinfo{author}{\bibfnamefont{B.}~\bibnamefont{Kubis}},
  \bibinfo{journal}{JHEP} \textbf{\bibinfo{volume}{08}}, \bibinfo{pages}{137}
  (\bibinfo{year}{2019}), \eprint{1907.01556}.

\bibitem[{\citenamefont{Davier et~al.}(2020)\citenamefont{Davier, Hoecker,
  Malaescu, and Zhang}}]{Davier:2019can}
\bibinfo{author}{\bibfnamefont{M.}~\bibnamefont{Davier}},
  \bibinfo{author}{\bibfnamefont{A.}~\bibnamefont{Hoecker}},
  \bibinfo{author}{\bibfnamefont{B.}~\bibnamefont{Malaescu}}, \bibnamefont{and}
  \bibinfo{author}{\bibfnamefont{Z.}~\bibnamefont{Zhang}},
  \bibinfo{journal}{Eur. Phys. J. C} \textbf{\bibinfo{volume}{80}},
  \bibinfo{pages}{241} (\bibinfo{year}{2020}), \bibinfo{note}{[Erratum: Eur.
  Phys. J. C {\bf 80}, 410 (2020)]}, \eprint{1908.00921}.

\bibitem[{\citenamefont{Keshavarzi
  et~al.}(2020{\natexlab{a}})\citenamefont{Keshavarzi, Nomura, and
  Teubner}}]{Keshavarzi:2019abf}
\bibinfo{author}{\bibfnamefont{A.}~\bibnamefont{Keshavarzi}},
  \bibinfo{author}{\bibfnamefont{D.}~\bibnamefont{Nomura}}, \bibnamefont{and}
  \bibinfo{author}{\bibfnamefont{T.}~\bibnamefont{Teubner}},
  \bibinfo{journal}{Phys. Rev. D} \textbf{\bibinfo{volume}{101}},
  \bibinfo{pages}{014029} (\bibinfo{year}{2020}{\natexlab{a}}),
  \eprint{1911.00367}.

\bibitem[{\citenamefont{Hoid et~al.}(2020)\citenamefont{Hoid, Hoferichter, and
  Kubis}}]{Hoid:2020xjs}
\bibinfo{author}{\bibfnamefont{B.-L.} \bibnamefont{Hoid}},
  \bibinfo{author}{\bibfnamefont{M.}~\bibnamefont{Hoferichter}},
  \bibnamefont{and} \bibinfo{author}{\bibfnamefont{B.}~\bibnamefont{Kubis}},
  \bibinfo{journal}{Eur. Phys. J. C} \textbf{\bibinfo{volume}{80}},
  \bibinfo{pages}{988} (\bibinfo{year}{2020}), \eprint{2007.12696}.

\bibitem[{\citenamefont{Kurz et~al.}(2014{\natexlab{a}})\citenamefont{Kurz,
  Liu, Marquard, and Steinhauser}}]{Kurz:2014wya}
\bibinfo{author}{\bibfnamefont{A.}~\bibnamefont{Kurz}},
  \bibinfo{author}{\bibfnamefont{T.}~\bibnamefont{Liu}},
  \bibinfo{author}{\bibfnamefont{P.}~\bibnamefont{Marquard}}, \bibnamefont{and}
  \bibinfo{author}{\bibfnamefont{M.}~\bibnamefont{Steinhauser}},
  \bibinfo{journal}{Phys. Lett. B} \textbf{\bibinfo{volume}{734}},
  \bibinfo{pages}{144} (\bibinfo{year}{2014}{\natexlab{a}}),
  \eprint{1403.6400}.

\bibitem[{\citenamefont{Melnikov and Vainshtein}(2004)}]{Melnikov:2003xd}
\bibinfo{author}{\bibfnamefont{K.}~\bibnamefont{Melnikov}} \bibnamefont{and}
  \bibinfo{author}{\bibfnamefont{A.}~\bibnamefont{Vainshtein}},
  \bibinfo{journal}{Phys. Rev. D} \textbf{\bibinfo{volume}{70}},
  \bibinfo{pages}{113006} (\bibinfo{year}{2004}), \eprint{hep-ph/0312226}.

\bibitem[{\citenamefont{Colangelo
  et~al.}(2014{\natexlab{a}})\citenamefont{Colangelo, Hoferichter, Procura, and
  Stoffer}}]{Colangelo:2014dfa}
\bibinfo{author}{\bibfnamefont{G.}~\bibnamefont{Colangelo}},
  \bibinfo{author}{\bibfnamefont{M.}~\bibnamefont{Hoferichter}},
  \bibinfo{author}{\bibfnamefont{M.}~\bibnamefont{Procura}}, \bibnamefont{and}
  \bibinfo{author}{\bibfnamefont{P.}~\bibnamefont{Stoffer}},
  \bibinfo{journal}{JHEP} \textbf{\bibinfo{volume}{09}}, \bibinfo{pages}{091}
  (\bibinfo{year}{2014}{\natexlab{a}}), \eprint{1402.7081}.

\bibitem[{\citenamefont{Colangelo
  et~al.}(2014{\natexlab{b}})\citenamefont{Colangelo, Hoferichter, Kubis,
  Procura, and Stoffer}}]{Colangelo:2014pva}
\bibinfo{author}{\bibfnamefont{G.}~\bibnamefont{Colangelo}},
  \bibinfo{author}{\bibfnamefont{M.}~\bibnamefont{Hoferichter}},
  \bibinfo{author}{\bibfnamefont{B.}~\bibnamefont{Kubis}},
  \bibinfo{author}{\bibfnamefont{M.}~\bibnamefont{Procura}}, \bibnamefont{and}
  \bibinfo{author}{\bibfnamefont{P.}~\bibnamefont{Stoffer}},
  \bibinfo{journal}{Phys. Lett. B} \textbf{\bibinfo{volume}{738}},
  \bibinfo{pages}{6} (\bibinfo{year}{2014}{\natexlab{b}}), \eprint{1408.2517}.

\bibitem[{\citenamefont{Colangelo et~al.}(2015)\citenamefont{Colangelo,
  Hoferichter, Procura, and Stoffer}}]{Colangelo:2015ama}
\bibinfo{author}{\bibfnamefont{G.}~\bibnamefont{Colangelo}},
  \bibinfo{author}{\bibfnamefont{M.}~\bibnamefont{Hoferichter}},
  \bibinfo{author}{\bibfnamefont{M.}~\bibnamefont{Procura}}, \bibnamefont{and}
  \bibinfo{author}{\bibfnamefont{P.}~\bibnamefont{Stoffer}},
  \bibinfo{journal}{JHEP} \textbf{\bibinfo{volume}{09}}, \bibinfo{pages}{074}
  (\bibinfo{year}{2015}), \eprint{1506.01386}.

\bibitem[{\citenamefont{Masjuan and
  S{\'a}nchez-Puertas}(2017)}]{Masjuan:2017tvw}
\bibinfo{author}{\bibfnamefont{P.}~\bibnamefont{Masjuan}} \bibnamefont{and}
  \bibinfo{author}{\bibfnamefont{P.}~\bibnamefont{S{\'a}nchez-Puertas}},
  \bibinfo{journal}{Phys. Rev. D} \textbf{\bibinfo{volume}{95}},
  \bibinfo{pages}{054026} (\bibinfo{year}{2017}), \eprint{1701.05829}.

\bibitem[{\citenamefont{Colangelo
  et~al.}(2017{\natexlab{a}})\citenamefont{Colangelo, Hoferichter, Procura, and
  Stoffer}}]{Colangelo:2017qdm}
\bibinfo{author}{\bibfnamefont{G.}~\bibnamefont{Colangelo}},
  \bibinfo{author}{\bibfnamefont{M.}~\bibnamefont{Hoferichter}},
  \bibinfo{author}{\bibfnamefont{M.}~\bibnamefont{Procura}}, \bibnamefont{and}
  \bibinfo{author}{\bibfnamefont{P.}~\bibnamefont{Stoffer}},
  \bibinfo{journal}{Phys. Rev. Lett.} \textbf{\bibinfo{volume}{118}},
  \bibinfo{pages}{232001} (\bibinfo{year}{2017}{\natexlab{a}}),
  \eprint{1701.06554}.

\bibitem[{\citenamefont{Colangelo
  et~al.}(2017{\natexlab{b}})\citenamefont{Colangelo, Hoferichter, Procura, and
  Stoffer}}]{Colangelo:2017fiz}
\bibinfo{author}{\bibfnamefont{G.}~\bibnamefont{Colangelo}},
  \bibinfo{author}{\bibfnamefont{M.}~\bibnamefont{Hoferichter}},
  \bibinfo{author}{\bibfnamefont{M.}~\bibnamefont{Procura}}, \bibnamefont{and}
  \bibinfo{author}{\bibfnamefont{P.}~\bibnamefont{Stoffer}},
  \bibinfo{journal}{JHEP} \textbf{\bibinfo{volume}{04}}, \bibinfo{pages}{161}
  (\bibinfo{year}{2017}{\natexlab{b}}), \eprint{1702.07347}.

\bibitem[{\citenamefont{Hoferichter
  et~al.}(2018{\natexlab{a}})\citenamefont{Hoferichter, Hoid, Kubis, Leupold,
  and Schneider}}]{Hoferichter:2018dmo}
\bibinfo{author}{\bibfnamefont{M.}~\bibnamefont{Hoferichter}},
  \bibinfo{author}{\bibfnamefont{B.-L.} \bibnamefont{Hoid}},
  \bibinfo{author}{\bibfnamefont{B.}~\bibnamefont{Kubis}},
  \bibinfo{author}{\bibfnamefont{S.}~\bibnamefont{Leupold}}, \bibnamefont{and}
  \bibinfo{author}{\bibfnamefont{S.~P.} \bibnamefont{Schneider}},
  \bibinfo{journal}{Phys. Rev. Lett.} \textbf{\bibinfo{volume}{121}},
  \bibinfo{pages}{112002} (\bibinfo{year}{2018}{\natexlab{a}}),
  \eprint{1805.01471}.

\bibitem[{\citenamefont{Hoferichter
  et~al.}(2018{\natexlab{b}})\citenamefont{Hoferichter, Hoid, Kubis, Leupold,
  and Schneider}}]{Hoferichter:2018kwz}
\bibinfo{author}{\bibfnamefont{M.}~\bibnamefont{Hoferichter}},
  \bibinfo{author}{\bibfnamefont{B.-L.} \bibnamefont{Hoid}},
  \bibinfo{author}{\bibfnamefont{B.}~\bibnamefont{Kubis}},
  \bibinfo{author}{\bibfnamefont{S.}~\bibnamefont{Leupold}}, \bibnamefont{and}
  \bibinfo{author}{\bibfnamefont{S.~P.} \bibnamefont{Schneider}},
  \bibinfo{journal}{JHEP} \textbf{\bibinfo{volume}{10}}, \bibinfo{pages}{141}
  (\bibinfo{year}{2018}{\natexlab{b}}), \eprint{1808.04823}.

\bibitem[{\citenamefont{G{\'e}rardin et~al.}(2019)\citenamefont{G{\'e}rardin,
  Meyer, and Nyffeler}}]{Gerardin:2019vio}
\bibinfo{author}{\bibfnamefont{A.}~\bibnamefont{G{\'e}rardin}},
  \bibinfo{author}{\bibfnamefont{H.~B.} \bibnamefont{Meyer}}, \bibnamefont{and}
  \bibinfo{author}{\bibfnamefont{A.}~\bibnamefont{Nyffeler}},
  \bibinfo{journal}{Phys. Rev. D} \textbf{\bibinfo{volume}{100}},
  \bibinfo{pages}{034520} (\bibinfo{year}{2019}), \eprint{1903.09471}.

\bibitem[{\citenamefont{Bijnens et~al.}(2019)\citenamefont{Bijnens,
  Hermansson-Truedsson, and Rodr{\'i}guez-S{\'a}nchez}}]{Bijnens:2019ghy}
\bibinfo{author}{\bibfnamefont{J.}~\bibnamefont{Bijnens}},
  \bibinfo{author}{\bibfnamefont{N.}~\bibnamefont{Hermansson-Truedsson}},
  \bibnamefont{and}
  \bibinfo{author}{\bibfnamefont{A.}~\bibnamefont{Rodr{\'i}guez-S{\'a}nchez}},
  \bibinfo{journal}{Phys. Lett. B} \textbf{\bibinfo{volume}{798}},
  \bibinfo{pages}{134994} (\bibinfo{year}{2019}), \eprint{1908.03331}.

\bibitem[{\citenamefont{Colangelo
  et~al.}(2020{\natexlab{a}})\citenamefont{Colangelo, Hagelstein, Hoferichter,
  Laub, and Stoffer}}]{Colangelo:2019lpu}
\bibinfo{author}{\bibfnamefont{G.}~\bibnamefont{Colangelo}},
  \bibinfo{author}{\bibfnamefont{F.}~\bibnamefont{Hagelstein}},
  \bibinfo{author}{\bibfnamefont{M.}~\bibnamefont{Hoferichter}},
  \bibinfo{author}{\bibfnamefont{L.}~\bibnamefont{Laub}}, \bibnamefont{and}
  \bibinfo{author}{\bibfnamefont{P.}~\bibnamefont{Stoffer}},
  \bibinfo{journal}{Phys. Rev. D} \textbf{\bibinfo{volume}{101}},
  \bibinfo{pages}{051501} (\bibinfo{year}{2020}{\natexlab{a}}),
  \eprint{1910.11881}.

\bibitem[{\citenamefont{Colangelo
  et~al.}(2020{\natexlab{b}})\citenamefont{Colangelo, Hagelstein, Hoferichter,
  Laub, and Stoffer}}]{Colangelo:2019uex}
\bibinfo{author}{\bibfnamefont{G.}~\bibnamefont{Colangelo}},
  \bibinfo{author}{\bibfnamefont{F.}~\bibnamefont{Hagelstein}},
  \bibinfo{author}{\bibfnamefont{M.}~\bibnamefont{Hoferichter}},
  \bibinfo{author}{\bibfnamefont{L.}~\bibnamefont{Laub}}, \bibnamefont{and}
  \bibinfo{author}{\bibfnamefont{P.}~\bibnamefont{Stoffer}},
  \bibinfo{journal}{JHEP} \textbf{\bibinfo{volume}{03}}, \bibinfo{pages}{101}
  (\bibinfo{year}{2020}{\natexlab{b}}), \eprint{1910.13432}.

\bibitem[{\citenamefont{Blum et~al.}(2020)\citenamefont{Blum, Christ, Hayakawa,
  Izubuchi, Jin, Jung, and Lehner}}]{Blum:2019ugy}
\bibinfo{author}{\bibfnamefont{T.}~\bibnamefont{Blum}},
  \bibinfo{author}{\bibfnamefont{N.}~\bibnamefont{Christ}},
  \bibinfo{author}{\bibfnamefont{M.}~\bibnamefont{Hayakawa}},
  \bibinfo{author}{\bibfnamefont{T.}~\bibnamefont{Izubuchi}},
  \bibinfo{author}{\bibfnamefont{L.}~\bibnamefont{Jin}},
  \bibinfo{author}{\bibfnamefont{C.}~\bibnamefont{Jung}}, \bibnamefont{and}
  \bibinfo{author}{\bibfnamefont{C.}~\bibnamefont{Lehner}},
  \bibinfo{journal}{Phys. Rev. Lett.} \textbf{\bibinfo{volume}{124}},
  \bibinfo{pages}{132002} (\bibinfo{year}{2020}), \eprint{1911.08123}.

\bibitem[{\citenamefont{Colangelo
  et~al.}(2014{\natexlab{c}})\citenamefont{Colangelo, Hoferichter, Nyffeler,
  Passera, and Stoffer}}]{Colangelo:2014qya}
\bibinfo{author}{\bibfnamefont{G.}~\bibnamefont{Colangelo}},
  \bibinfo{author}{\bibfnamefont{M.}~\bibnamefont{Hoferichter}},
  \bibinfo{author}{\bibfnamefont{A.}~\bibnamefont{Nyffeler}},
  \bibinfo{author}{\bibfnamefont{M.}~\bibnamefont{Passera}}, \bibnamefont{and}
  \bibinfo{author}{\bibfnamefont{P.}~\bibnamefont{Stoffer}},
  \bibinfo{journal}{Phys. Lett. B} \textbf{\bibinfo{volume}{735}},
  \bibinfo{pages}{90} (\bibinfo{year}{2014}{\natexlab{c}}), \eprint{1403.7512}.

\bibitem[{\citenamefont{Calmet et~al.}(1976)\citenamefont{Calmet, Narison,
  Perrottet, and de~Rafael}}]{Calmet:1976kd}
\bibinfo{author}{\bibfnamefont{J.}~\bibnamefont{Calmet}},
  \bibinfo{author}{\bibfnamefont{S.}~\bibnamefont{Narison}},
  \bibinfo{author}{\bibfnamefont{M.}~\bibnamefont{Perrottet}},
  \bibnamefont{and}
  \bibinfo{author}{\bibfnamefont{E.}~\bibnamefont{de~Rafael}},
  \bibinfo{journal}{Phys. Lett. B} \textbf{\bibinfo{volume}{61}},
  \bibinfo{pages}{283} (\bibinfo{year}{1976}).

\bibitem[{\citenamefont{Pauk and Vanderhaeghen}(2014)}]{Pauk:2014rta}
\bibinfo{author}{\bibfnamefont{V.}~\bibnamefont{Pauk}} \bibnamefont{and}
  \bibinfo{author}{\bibfnamefont{M.}~\bibnamefont{Vanderhaeghen}},
  \bibinfo{journal}{Eur. Phys. J. C} \textbf{\bibinfo{volume}{74}},
  \bibinfo{pages}{3008} (\bibinfo{year}{2014}), \eprint{1401.0832}.

\bibitem[{\citenamefont{Danilkin and Vanderhaeghen}(2017)}]{Danilkin:2016hnh}
\bibinfo{author}{\bibfnamefont{I.}~\bibnamefont{Danilkin}} \bibnamefont{and}
  \bibinfo{author}{\bibfnamefont{M.}~\bibnamefont{Vanderhaeghen}},
  \bibinfo{journal}{Phys. Rev. D} \textbf{\bibinfo{volume}{95}},
  \bibinfo{pages}{014019} (\bibinfo{year}{2017}), \eprint{1611.04646}.

\bibitem[{\citenamefont{Jegerlehner}(2017)}]{Jegerlehner:2017gek}
\bibinfo{author}{\bibfnamefont{F.}~\bibnamefont{Jegerlehner}},
  \bibinfo{journal}{Springer Tracts Mod. Phys.} \textbf{\bibinfo{volume}{274}},
  \bibinfo{pages}{1} (\bibinfo{year}{2017}).

\bibitem[{\citenamefont{Knecht et~al.}(2018)\citenamefont{Knecht, Narison,
  Rabemananjara, and Rabetiarivony}}]{Knecht:2018sci}
\bibinfo{author}{\bibfnamefont{M.}~\bibnamefont{Knecht}},
  \bibinfo{author}{\bibfnamefont{S.}~\bibnamefont{Narison}},
  \bibinfo{author}{\bibfnamefont{A.}~\bibnamefont{Rabemananjara}},
  \bibnamefont{and}
  \bibinfo{author}{\bibfnamefont{D.}~\bibnamefont{Rabetiarivony}},
  \bibinfo{journal}{Phys. Lett. B} \textbf{\bibinfo{volume}{787}},
  \bibinfo{pages}{111} (\bibinfo{year}{2018}), \eprint{1808.03848}.

\bibitem[{\citenamefont{Eichmann et~al.}(2020)\citenamefont{Eichmann, Fischer,
  and Williams}}]{Eichmann:2019bqf}
\bibinfo{author}{\bibfnamefont{G.}~\bibnamefont{Eichmann}},
  \bibinfo{author}{\bibfnamefont{C.~S.} \bibnamefont{Fischer}},
  \bibnamefont{and} \bibinfo{author}{\bibfnamefont{R.}~\bibnamefont{Williams}},
  \bibinfo{journal}{Phys. Rev. D} \textbf{\bibinfo{volume}{101}},
  \bibinfo{pages}{054015} (\bibinfo{year}{2020}), \eprint{1910.06795}.

\bibitem[{\citenamefont{Roig and S{\'a}nchez-Puertas}(2020)}]{Roig:2019reh}
\bibinfo{author}{\bibfnamefont{P.}~\bibnamefont{Roig}} \bibnamefont{and}
  \bibinfo{author}{\bibfnamefont{P.}~\bibnamefont{S{\'a}nchez-Puertas}},
  \bibinfo{journal}{Phys. Rev. D} \textbf{\bibinfo{volume}{101}},
  \bibinfo{pages}{074019} (\bibinfo{year}{2020}), \eprint{1910.02881}.

\bibitem[{\citenamefont{Borsanyi et~al.}(2021)}]{Borsanyi:2020mff}
\bibinfo{author}{\bibfnamefont{S.}~\bibnamefont{Borsanyi}}
  \bibnamefont{et~al.}, \bibinfo{journal}{Nature}
  \textbf{\bibinfo{volume}{593}}, \bibinfo{pages}{51} (\bibinfo{year}{2021}),
  \eprint{2002.12347}.

\bibitem[{\citenamefont{Lehner and Meyer}(2020)}]{Lehner:2020crt}
\bibinfo{author}{\bibfnamefont{C.}~\bibnamefont{Lehner}} \bibnamefont{and}
  \bibinfo{author}{\bibfnamefont{A.~S.} \bibnamefont{Meyer}},
  \bibinfo{journal}{Phys. Rev. D} \textbf{\bibinfo{volume}{101}},
  \bibinfo{pages}{074515} (\bibinfo{year}{2020}), \eprint{2003.04177}.

\bibitem[{\citenamefont{Crivellin et~al.}(2020)\citenamefont{Crivellin,
  Hoferichter, Manzari, and Montull}}]{Crivellin:2020zul}
\bibinfo{author}{\bibfnamefont{A.}~\bibnamefont{Crivellin}},
  \bibinfo{author}{\bibfnamefont{M.}~\bibnamefont{Hoferichter}},
  \bibinfo{author}{\bibfnamefont{C.~A.} \bibnamefont{Manzari}},
  \bibnamefont{and} \bibinfo{author}{\bibfnamefont{M.}~\bibnamefont{Montull}},
  \bibinfo{journal}{Phys. Rev. Lett.} \textbf{\bibinfo{volume}{125}},
  \bibinfo{pages}{091801} (\bibinfo{year}{2020}), \eprint{2003.04886}.

\bibitem[{\citenamefont{Keshavarzi
  et~al.}(2020{\natexlab{b}})\citenamefont{Keshavarzi, Marciano, Passera, and
  Sirlin}}]{Keshavarzi:2020bfy}
\bibinfo{author}{\bibfnamefont{A.}~\bibnamefont{Keshavarzi}},
  \bibinfo{author}{\bibfnamefont{W.~J.} \bibnamefont{Marciano}},
  \bibinfo{author}{\bibfnamefont{M.}~\bibnamefont{Passera}}, \bibnamefont{and}
  \bibinfo{author}{\bibfnamefont{A.}~\bibnamefont{Sirlin}},
  \bibinfo{journal}{Phys. Rev. D} \textbf{\bibinfo{volume}{102}},
  \bibinfo{pages}{033002} (\bibinfo{year}{2020}{\natexlab{b}}),
  \eprint{2006.12666}.

\bibitem[{\citenamefont{Malaescu and Schott}(2021)}]{Malaescu:2020zuc}
\bibinfo{author}{\bibfnamefont{B.}~\bibnamefont{Malaescu}} \bibnamefont{and}
  \bibinfo{author}{\bibfnamefont{M.}~\bibnamefont{Schott}},
  \bibinfo{journal}{Eur. Phys. J. C} \textbf{\bibinfo{volume}{81}},
  \bibinfo{pages}{46} (\bibinfo{year}{2021}), \eprint{2008.08107}.

\bibitem[{\citenamefont{Colangelo
  et~al.}(2021{\natexlab{a}})\citenamefont{Colangelo, Hoferichter, and
  Stoffer}}]{Colangelo:2020lcg}
\bibinfo{author}{\bibfnamefont{G.}~\bibnamefont{Colangelo}},
  \bibinfo{author}{\bibfnamefont{M.}~\bibnamefont{Hoferichter}},
  \bibnamefont{and} \bibinfo{author}{\bibfnamefont{P.}~\bibnamefont{Stoffer}},
  \bibinfo{journal}{Phys. Lett. B} \textbf{\bibinfo{volume}{814}},
  \bibinfo{pages}{136073} (\bibinfo{year}{2021}{\natexlab{a}}),
  \eprint{2010.07943}.

\bibitem[{\citenamefont{Hoferichter and Stoffer}(2020)}]{Hoferichter:2020lap}
\bibinfo{author}{\bibfnamefont{M.}~\bibnamefont{Hoferichter}} \bibnamefont{and}
  \bibinfo{author}{\bibfnamefont{P.}~\bibnamefont{Stoffer}},
  \bibinfo{journal}{JHEP} \textbf{\bibinfo{volume}{05}}, \bibinfo{pages}{159}
  (\bibinfo{year}{2020}), \eprint{2004.06127}.

\bibitem[{\citenamefont{L\"udtke and Procura}(2020)}]{Ludtke:2020moa}
\bibinfo{author}{\bibfnamefont{J.}~\bibnamefont{L\"udtke}} \bibnamefont{and}
  \bibinfo{author}{\bibfnamefont{M.}~\bibnamefont{Procura}},
  \bibinfo{journal}{Eur. Phys. J. C} \textbf{\bibinfo{volume}{80}},
  \bibinfo{pages}{1108} (\bibinfo{year}{2020}), \eprint{2006.00007}.

\bibitem[{\citenamefont{Bijnens et~al.}(2020)\citenamefont{Bijnens,
  Hermansson-Truedsson, Laub, and Rodr{\'i}guez-S\'anchez}}]{Bijnens:2020xnl}
\bibinfo{author}{\bibfnamefont{J.}~\bibnamefont{Bijnens}},
  \bibinfo{author}{\bibfnamefont{N.}~\bibnamefont{Hermansson-Truedsson}},
  \bibinfo{author}{\bibfnamefont{L.}~\bibnamefont{Laub}}, \bibnamefont{and}
  \bibinfo{author}{\bibfnamefont{A.}~\bibnamefont{Rodr{\'i}guez-S\'anchez}},
  \bibinfo{journal}{JHEP} \textbf{\bibinfo{volume}{10}}, \bibinfo{pages}{203}
  (\bibinfo{year}{2020}), \eprint{2008.13487}.

\bibitem[{\citenamefont{Bijnens et~al.}(2021)\citenamefont{Bijnens,
  Hermansson-Truedsson, Laub, and Rodr\'iguez-S\'anchez}}]{Bijnens:2021jqo}
\bibinfo{author}{\bibfnamefont{J.}~\bibnamefont{Bijnens}},
  \bibinfo{author}{\bibfnamefont{N.}~\bibnamefont{Hermansson-Truedsson}},
  \bibinfo{author}{\bibfnamefont{L.}~\bibnamefont{Laub}}, \bibnamefont{and}
  \bibinfo{author}{\bibfnamefont{A.}~\bibnamefont{Rodr\'iguez-S\'anchez}},
  \bibinfo{journal}{JHEP} \textbf{\bibinfo{volume}{04}}, \bibinfo{pages}{240}
  (\bibinfo{year}{2021}), \eprint{2101.09169}.

\bibitem[{\citenamefont{Zanke et~al.}(2021)\citenamefont{Zanke, Hoferichter,
  and Kubis}}]{Zanke:2021wiq}
\bibinfo{author}{\bibfnamefont{M.}~\bibnamefont{Zanke}},
  \bibinfo{author}{\bibfnamefont{M.}~\bibnamefont{Hoferichter}},
  \bibnamefont{and} \bibinfo{author}{\bibfnamefont{B.}~\bibnamefont{Kubis}},
  \bibinfo{journal}{JHEP} \textbf{\bibinfo{volume}{07}}, \bibinfo{pages}{106}
  (\bibinfo{year}{2021}), \eprint{2103.09829}.

\bibitem[{\citenamefont{Chao et~al.}(2021)\citenamefont{Chao, Hudspith,
  G\'erardin, Green, Meyer, and Ottnad}}]{Chao:2021tvp}
\bibinfo{author}{\bibfnamefont{E.-H.} \bibnamefont{Chao}},
  \bibinfo{author}{\bibfnamefont{R.~J.} \bibnamefont{Hudspith}},
  \bibinfo{author}{\bibfnamefont{A.}~\bibnamefont{G\'erardin}},
  \bibinfo{author}{\bibfnamefont{J.~R.} \bibnamefont{Green}},
  \bibinfo{author}{\bibfnamefont{H.~B.} \bibnamefont{Meyer}}, \bibnamefont{and}
  \bibinfo{author}{\bibfnamefont{K.}~\bibnamefont{Ottnad}},
  \bibinfo{journal}{Eur. Phys. J. C} \textbf{\bibinfo{volume}{81}},
  \bibinfo{pages}{651} (\bibinfo{year}{2021}), \eprint{2104.02632}.

\bibitem[{\citenamefont{Danilkin et~al.}(2021)\citenamefont{Danilkin,
  Hoferichter, and Stoffer}}]{Danilkin:2021icn}
\bibinfo{author}{\bibfnamefont{I.}~\bibnamefont{Danilkin}},
  \bibinfo{author}{\bibfnamefont{M.}~\bibnamefont{Hoferichter}},
  \bibnamefont{and} \bibinfo{author}{\bibfnamefont{P.}~\bibnamefont{Stoffer}},
  \bibinfo{journal}{Phys. Lett. B} \textbf{\bibinfo{volume}{820}},
  \bibinfo{pages}{136502} (\bibinfo{year}{2021}), \eprint{2105.01666}.

\bibitem[{\citenamefont{Colangelo
  et~al.}(2021{\natexlab{b}})\citenamefont{Colangelo, Hagelstein, Hoferichter,
  Laub, and Stoffer}}]{Colangelo:2021nkr}
\bibinfo{author}{\bibfnamefont{G.}~\bibnamefont{Colangelo}},
  \bibinfo{author}{\bibfnamefont{F.}~\bibnamefont{Hagelstein}},
  \bibinfo{author}{\bibfnamefont{M.}~\bibnamefont{Hoferichter}},
  \bibinfo{author}{\bibfnamefont{L.}~\bibnamefont{Laub}}, \bibnamefont{and}
  \bibinfo{author}{\bibfnamefont{P.}~\bibnamefont{Stoffer}},
  \bibinfo{journal}{Eur. Phys. J. C} \textbf{\bibinfo{volume}{81}},
  \bibinfo{pages}{702} (\bibinfo{year}{2021}{\natexlab{b}}),
  \eprint{2106.13222}.

\bibitem[{\citenamefont{Bennett et~al.}(2006)}]{bennett:2006fi}
\bibinfo{author}{\bibfnamefont{G.~W.} \bibnamefont{Bennett}}
  \bibnamefont{et~al.} (\bibinfo{collaboration}{Muon $g-2$}),
  \bibinfo{journal}{Phys. Rev. D} \textbf{\bibinfo{volume}{73}},
  \bibinfo{pages}{072003} (\bibinfo{year}{2006}), \eprint{hep-ex/0602035}.

\bibitem[{\citenamefont{Abi et~al.}(2021)}]{Abi:2021gix}
\bibinfo{author}{\bibfnamefont{B.}~\bibnamefont{Abi}} \bibnamefont{et~al.}
  (\bibinfo{collaboration}{Muon $g-2$}), \bibinfo{journal}{Phys. Rev. Lett.}
  \textbf{\bibinfo{volume}{126}}, \bibinfo{pages}{141801}
  (\bibinfo{year}{2021}), \eprint{2104.03281}.

\bibitem[{\citenamefont{Albahri et~al.}(2021{\natexlab{a}})}]{Albahri:2021ixb}
\bibinfo{author}{\bibfnamefont{T.}~\bibnamefont{Albahri}} \bibnamefont{et~al.}
  (\bibinfo{collaboration}{Muon $g-2$}), \bibinfo{journal}{Phys. Rev. D}
  \textbf{\bibinfo{volume}{103}}, \bibinfo{pages}{072002}
  (\bibinfo{year}{2021}{\natexlab{a}}), \eprint{2104.03247}.

\bibitem[{\citenamefont{Albahri et~al.}(2021{\natexlab{b}})}]{Albahri:2021kmg}
\bibinfo{author}{\bibfnamefont{T.}~\bibnamefont{Albahri}} \bibnamefont{et~al.}
  (\bibinfo{collaboration}{Muon $g-2$}), \bibinfo{journal}{Phys. Rev. A}
  \textbf{\bibinfo{volume}{103}}, \bibinfo{pages}{042208}
  (\bibinfo{year}{2021}{\natexlab{b}}), \eprint{2104.03201}.

\bibitem[{\citenamefont{Albahri et~al.}(2021{\natexlab{c}})}]{Albahri:2021mtf}
\bibinfo{author}{\bibfnamefont{T.}~\bibnamefont{Albahri}} \bibnamefont{et~al.}
  (\bibinfo{collaboration}{Muon $g-2$}), \bibinfo{journal}{Phys. Rev. Accel.
  Beams} \textbf{\bibinfo{volume}{24}}, \bibinfo{pages}{044002}
  (\bibinfo{year}{2021}{\natexlab{c}}), \eprint{2104.03240}.

\bibitem[{\citenamefont{Grange et~al.}(2015)}]{Muong-2:2015xgu}
\bibinfo{author}{\bibfnamefont{J.}~\bibnamefont{Grange}} \bibnamefont{et~al.}
  (\bibinfo{collaboration}{Muon $g-2$}) (\bibinfo{year}{2015}),
  \eprint{1501.06858}.

\bibitem[{\citenamefont{Abe et~al.}(2019)}]{Abe:2019thb}
\bibinfo{author}{\bibfnamefont{M.}~\bibnamefont{Abe}} \bibnamefont{et~al.},
  \bibinfo{journal}{PTEP} \textbf{\bibinfo{volume}{2019}},
  \bibinfo{pages}{053C02} (\bibinfo{year}{2019}), \eprint{1901.03047}.

\bibitem[{\citenamefont{Aiba et~al.}(2021)}]{Aiba:2021bxe}
\bibinfo{author}{\bibfnamefont{M.}~\bibnamefont{Aiba}} \bibnamefont{et~al.}
  (\bibinfo{year}{2021}), \eprint{2111.05788}.

\bibitem[{\citenamefont{Teubner}(1995)}]{Teubner:1995}
\bibinfo{author}{\bibfnamefont{T.}~\bibnamefont{Teubner}}
  (\bibinfo{year}{1995}), \bibinfo{note}{{PhD thesis, Karlsruhe}}.

\bibitem[{\citenamefont{Hoang}(1995)}]{Hoang:1995}
\bibinfo{author}{\bibfnamefont{A.~H.} \bibnamefont{Hoang}}
  (\bibinfo{year}{1995}), \bibinfo{note}{{PhD thesis, Karlsruhe}}.

\bibitem[{\citenamefont{Hoang et~al.}(1994{\natexlab{a}})\citenamefont{Hoang,
  Je{\.{z}}abek, K{\"u}hn, and Teubner}}]{Hoang:1993ns}
\bibinfo{author}{\bibfnamefont{A.~H.} \bibnamefont{Hoang}},
  \bibinfo{author}{\bibfnamefont{M.}~\bibnamefont{Je{\.{z}}abek}},
  \bibinfo{author}{\bibfnamefont{J.~H.} \bibnamefont{K{\"u}hn}},
  \bibnamefont{and} \bibinfo{author}{\bibfnamefont{T.}~\bibnamefont{Teubner}},
  \bibinfo{journal}{Phys. Lett. B} \textbf{\bibinfo{volume}{325}},
  \bibinfo{pages}{495} (\bibinfo{year}{1994}{\natexlab{a}}),
  \bibinfo{note}{[Erratum: Phys. Lett. B {\bf 327}, 439 (1994)]},
  \eprint{hep-ph/9401283}.

\bibitem[{\citenamefont{Hoang et~al.}(1994{\natexlab{b}})\citenamefont{Hoang,
  Je{\.{z}}abek, K{\"u}hn, and Teubner}}]{Hoang:1994it}
\bibinfo{author}{\bibfnamefont{A.~H.} \bibnamefont{Hoang}},
  \bibinfo{author}{\bibfnamefont{M.}~\bibnamefont{Je{\.{z}}abek}},
  \bibinfo{author}{\bibfnamefont{J.~H.} \bibnamefont{K{\"u}hn}},
  \bibnamefont{and} \bibinfo{author}{\bibfnamefont{T.}~\bibnamefont{Teubner}},
  \bibinfo{journal}{Phys. Lett. B} \textbf{\bibinfo{volume}{338}},
  \bibinfo{pages}{330} (\bibinfo{year}{1994}{\natexlab{b}}),
  \eprint{hep-ph/9407338}.

\bibitem[{\citenamefont{Hoang et~al.}(1995{\natexlab{a}})\citenamefont{Hoang,
  K{\"u}hn, and Teubner}}]{Hoang:1995ex}
\bibinfo{author}{\bibfnamefont{A.~H.} \bibnamefont{Hoang}},
  \bibinfo{author}{\bibfnamefont{J.~H.} \bibnamefont{K{\"u}hn}},
  \bibnamefont{and} \bibinfo{author}{\bibfnamefont{T.}~\bibnamefont{Teubner}},
  \bibinfo{journal}{Nucl. Phys. B} \textbf{\bibinfo{volume}{452}},
  \bibinfo{pages}{173} (\bibinfo{year}{1995}{\natexlab{a}}),
  \eprint{hep-ph/9505262}.

\bibitem[{\citenamefont{Hoang et~al.}(1995{\natexlab{b}})\citenamefont{Hoang,
  K{\"u}hn, and Teubner}}]{Hoang:1995ht}
\bibinfo{author}{\bibfnamefont{A.~H.} \bibnamefont{Hoang}},
  \bibinfo{author}{\bibfnamefont{J.~H.} \bibnamefont{K{\"u}hn}},
  \bibnamefont{and} \bibinfo{author}{\bibfnamefont{T.}~\bibnamefont{Teubner}},
  \bibinfo{journal}{Nucl. Phys. B} \textbf{\bibinfo{volume}{455}},
  \bibinfo{pages}{3} (\bibinfo{year}{1995}{\natexlab{b}}),
  \eprint{hep-ph/9507255}.

\bibitem[{\citenamefont{Brodsky et~al.}(1995)\citenamefont{Brodsky, Hoang,
  K{\"u}hn, and Teubner}}]{Brodsky:1995ds}
\bibinfo{author}{\bibfnamefont{S.~J.} \bibnamefont{Brodsky}},
  \bibinfo{author}{\bibfnamefont{A.~H.} \bibnamefont{Hoang}},
  \bibinfo{author}{\bibfnamefont{J.~H.} \bibnamefont{K{\"u}hn}},
  \bibnamefont{and} \bibinfo{author}{\bibfnamefont{T.}~\bibnamefont{Teubner}},
  \bibinfo{journal}{Phys. Lett. B} \textbf{\bibinfo{volume}{359}},
  \bibinfo{pages}{355} (\bibinfo{year}{1995}), \eprint{hep-ph/9508274}.

\bibitem[{\citenamefont{Chetyrkin et~al.}(1996)\citenamefont{Chetyrkin, Hoang,
  K{\"u}hn, Steinhauser, and Teubner}}]{Chetyrkin:1996yp}
\bibinfo{author}{\bibfnamefont{K.~G.} \bibnamefont{Chetyrkin}},
  \bibinfo{author}{\bibfnamefont{A.~H.} \bibnamefont{Hoang}},
  \bibinfo{author}{\bibfnamefont{J.~H.} \bibnamefont{K{\"u}hn}},
  \bibinfo{author}{\bibfnamefont{M.}~\bibnamefont{Steinhauser}},
  \bibnamefont{and} \bibinfo{author}{\bibfnamefont{T.}~\bibnamefont{Teubner}},
  \bibinfo{journal}{Phys. Lett. B} \textbf{\bibinfo{volume}{384}},
  \bibinfo{pages}{233} (\bibinfo{year}{1996}), \eprint{hep-ph/9603313}.

\bibitem[{\citenamefont{Hoang and Teubner}(1998)}]{Hoang:1997ca}
\bibinfo{author}{\bibfnamefont{A.~H.} \bibnamefont{Hoang}} \bibnamefont{and}
  \bibinfo{author}{\bibfnamefont{T.}~\bibnamefont{Teubner}},
  \bibinfo{journal}{Nucl. Phys. B} \textbf{\bibinfo{volume}{519}},
  \bibinfo{pages}{285} (\bibinfo{year}{1998}), \eprint{hep-ph/9707496}.

\bibitem[{app()}]{appendix}
\bibinfo{note}{{The appendix summarizes the QED spectral functions.}}

\bibitem[{\citenamefont{Nio}(2021)}]{Nio:2021}
\bibinfo{author}{\bibfnamefont{M.}~\bibnamefont{Nio}} (\bibinfo{year}{2021}),
  \bibinfo{note}{{private communication}}.

\bibitem[{\citenamefont{Laporta}(1993)}]{Laporta:1993ds}
\bibinfo{author}{\bibfnamefont{S.}~\bibnamefont{Laporta}},
  \bibinfo{journal}{Phys. Lett. B} \textbf{\bibinfo{volume}{312}},
  \bibinfo{pages}{495} (\bibinfo{year}{1993}), \eprint{hep-ph/9306324}.

\bibitem[{\citenamefont{Kinoshita and Nio}(2006)}]{Kinoshita:2005zr}
\bibinfo{author}{\bibfnamefont{T.}~\bibnamefont{Kinoshita}} \bibnamefont{and}
  \bibinfo{author}{\bibfnamefont{M.}~\bibnamefont{Nio}},
  \bibinfo{journal}{Phys. Rev. D} \textbf{\bibinfo{volume}{73}},
  \bibinfo{pages}{013003} (\bibinfo{year}{2006}), \eprint{hep-ph/0507249}.

\bibitem[{\citenamefont{Kataev}(2012)}]{Kataev:2012kn}
\bibinfo{author}{\bibfnamefont{A.~L.} \bibnamefont{Kataev}},
  \bibinfo{journal}{Phys. Rev. D} \textbf{\bibinfo{volume}{86}},
  \bibinfo{pages}{013010} (\bibinfo{year}{2012}), \eprint{1205.6191}.

\bibitem[{\citenamefont{Kurz et~al.}(2014{\natexlab{b}})\citenamefont{Kurz,
  Liu, Marquard, and Steinhauser}}]{Kurz:2013exa}
\bibinfo{author}{\bibfnamefont{A.}~\bibnamefont{Kurz}},
  \bibinfo{author}{\bibfnamefont{T.}~\bibnamefont{Liu}},
  \bibinfo{author}{\bibfnamefont{P.}~\bibnamefont{Marquard}}, \bibnamefont{and}
  \bibinfo{author}{\bibfnamefont{M.}~\bibnamefont{Steinhauser}},
  \bibinfo{journal}{Nucl. Phys. B} \textbf{\bibinfo{volume}{879}},
  \bibinfo{pages}{1} (\bibinfo{year}{2014}{\natexlab{b}}), \eprint{1311.2471}.

\bibitem[{\citenamefont{Kurz et~al.}(2015)\citenamefont{Kurz, Liu, Marquard,
  Smirnov, Smirnov, and Steinhauser}}]{Kurz:2015bia}
\bibinfo{author}{\bibfnamefont{A.}~\bibnamefont{Kurz}},
  \bibinfo{author}{\bibfnamefont{T.}~\bibnamefont{Liu}},
  \bibinfo{author}{\bibfnamefont{P.}~\bibnamefont{Marquard}},
  \bibinfo{author}{\bibfnamefont{A.~V.} \bibnamefont{Smirnov}},
  \bibinfo{author}{\bibfnamefont{V.~A.} \bibnamefont{Smirnov}},
  \bibnamefont{and}
  \bibinfo{author}{\bibfnamefont{M.}~\bibnamefont{Steinhauser}},
  \bibinfo{journal}{Phys. Rev. D} \textbf{\bibinfo{volume}{92}},
  \bibinfo{pages}{073019} (\bibinfo{year}{2015}), \eprint{1508.00901}.

\bibitem[{\citenamefont{Kurz et~al.}(2016)\citenamefont{Kurz, Liu, Marquard,
  Smirnov, Smirnov, and Steinhauser}}]{Kurz:2016bau}
\bibinfo{author}{\bibfnamefont{A.}~\bibnamefont{Kurz}},
  \bibinfo{author}{\bibfnamefont{T.}~\bibnamefont{Liu}},
  \bibinfo{author}{\bibfnamefont{P.}~\bibnamefont{Marquard}},
  \bibinfo{author}{\bibfnamefont{A.}~\bibnamefont{Smirnov}},
  \bibinfo{author}{\bibfnamefont{V.}~\bibnamefont{Smirnov}}, \bibnamefont{and}
  \bibinfo{author}{\bibfnamefont{M.}~\bibnamefont{Steinhauser}},
  \bibinfo{journal}{Phys. Rev. D} \textbf{\bibinfo{volume}{93}},
  \bibinfo{pages}{053017} (\bibinfo{year}{2016}), \eprint{1602.02785}.

\bibitem[{\citenamefont{Hoefer et~al.}(2002)\citenamefont{Hoefer, Gluza, and
  Jegerlehner}}]{Hoefer:2001mx}
\bibinfo{author}{\bibfnamefont{A.}~\bibnamefont{Hoefer}},
  \bibinfo{author}{\bibfnamefont{J.}~\bibnamefont{Gluza}}, \bibnamefont{and}
  \bibinfo{author}{\bibfnamefont{F.}~\bibnamefont{Jegerlehner}},
  \bibinfo{journal}{Eur. Phys. J. C} \textbf{\bibinfo{volume}{24}},
  \bibinfo{pages}{51} (\bibinfo{year}{2002}), \eprint{hep-ph/0107154}.

\bibitem[{\citenamefont{Czy{\.z} et~al.}(2005)\citenamefont{Czy{\.z},
  Grzeli{\'n}ska, K{\"u}hn, and Rodrigo}}]{Czyz:2004rj}
\bibinfo{author}{\bibfnamefont{H.}~\bibnamefont{Czy{\.z}}},
  \bibinfo{author}{\bibfnamefont{A.}~\bibnamefont{Grzeli{\'n}ska}},
  \bibinfo{author}{\bibfnamefont{J.~H.} \bibnamefont{K{\"u}hn}},
  \bibnamefont{and} \bibinfo{author}{\bibfnamefont{G.}~\bibnamefont{Rodrigo}},
  \bibinfo{journal}{Eur. Phys. J. C} \textbf{\bibinfo{volume}{39}},
  \bibinfo{pages}{411} (\bibinfo{year}{2005}), \eprint{hep-ph/0404078}.

\bibitem[{\citenamefont{Gluza et~al.}(2003)\citenamefont{Gluza, Hoefer, Jadach,
  and Jegerlehner}}]{Gluza:2002ui}
\bibinfo{author}{\bibfnamefont{J.}~\bibnamefont{Gluza}},
  \bibinfo{author}{\bibfnamefont{A.}~\bibnamefont{Hoefer}},
  \bibinfo{author}{\bibfnamefont{S.}~\bibnamefont{Jadach}}, \bibnamefont{and}
  \bibinfo{author}{\bibfnamefont{F.}~\bibnamefont{Jegerlehner}},
  \bibinfo{journal}{Eur. Phys. J. C} \textbf{\bibinfo{volume}{28}},
  \bibinfo{pages}{261} (\bibinfo{year}{2003}), \eprint{hep-ph/0212386}.

\bibitem[{\citenamefont{Bystritskiy et~al.}(2005)\citenamefont{Bystritskiy,
  Kuraev, Fedotovich, and Ignatov}}]{Bystritskiy:2005ib}
\bibinfo{author}{\bibfnamefont{Y.~M.} \bibnamefont{Bystritskiy}},
  \bibinfo{author}{\bibfnamefont{E.~A.} \bibnamefont{Kuraev}},
  \bibinfo{author}{\bibfnamefont{G.~V.} \bibnamefont{Fedotovich}},
  \bibnamefont{and} \bibinfo{author}{\bibfnamefont{F.~V.}
  \bibnamefont{Ignatov}}, \bibinfo{journal}{Phys. Rev. D}
  \textbf{\bibinfo{volume}{72}}, \bibinfo{pages}{114019}
  (\bibinfo{year}{2005}), \eprint{hep-ph/0505236}.

\bibitem[{\citenamefont{Abbiendi et~al.}(2022)}]{Abbiendi:2022liz}
\bibinfo{author}{\bibfnamefont{G.}~\bibnamefont{Abbiendi}} \bibnamefont{et~al.}
  (\bibinfo{year}{2022}), \eprint{2201.12102}.

\end{thebibliography}

\end{document}